\documentclass[lettersize,journal]{IEEEtran}
\usepackage{amsmath,amsfonts}
\usepackage{algorithmic}
\usepackage{algorithm}
\usepackage{array}
\usepackage[caption=false,font=normalsize,labelfont=sf,textfont=sf]{subfig}
\usepackage{textcomp}
\usepackage{stfloats}
\usepackage{url}
\usepackage{verbatim}
\usepackage{graphicx}
\usepackage{cite}
\begin{document}

\title{Temporal Attention Pooling for Frequency Dynamic Convolution in Sound Event Detection}

\author{Hyeonuk Nam,~\IEEEmembership{member,~IEEE,} Yong-Hwa Park,~\IEEEmembership{member,~IEEE,}
\thanks{This paper was produced by the IEEE Publication Technology Group. They are in Piscataway, NJ.}
\thanks{Manuscript received April 19, 2021; revised August 16, 2021.}}

\markboth{Journal of \LaTeX\ Class Files,~Vol.~14, No.~8, August~2021}%
{Shell \MakeLowercase{\textit{et al.}}: A Sample Article Using IEEEtran.cls for IEEE Journals}

\IEEEpubid{0000--0000/00\$00.00~\copyright~2025 IEEE}

\maketitle

\begin{abstract}
Recent advances in deep learning, particularly frequency dynamic convolution (FDY conv), have significantly improved sound event detection (SED) by enabling frequency-adaptive feature extraction. However, FDY conv relies on \textit{temporal average pooling}, which treats all temporal frames equally, limiting its ability to capture transient sound events such as alarm bells, door knocks, and speech plosives. To address this limitation, we propose \textit{temporal attention pooling frequency dynamic convolution (TFD conv)} to replace temporal average pooling with \textit{temporal attention pooling (TAP)}. TAP adaptively weights temporal features through three complementary mechanisms: \textit{time attention pooling (TA)} for emphasizing salient features, \textit{velocity attention pooling (VA)} for capturing transient changes, and conventional \textit{average pooling} for robustness to stationary signals. Ablation studies show that TFD conv improves average PSDS1 by 3.02\% over FDY conv with only a 14.8\% increase in parameter count. Classwise ANOVA and Tukey HSD analysis further demonstrate that TFD conv significantly enhances detection performance for transient-heavy events, outperforming existing FDY conv models. Notably, TFD conv achieves a maximum PSDS1 score of 0.456, surpassing previous state-of-the-art SED systems. We also explore the compatibility of TAP with other FDY conv variants, including dilated FDY conv (DFD conv), partial FDY conv (PFD conv), and multi-dilated FDY conv (MDFD conv). Among these, the integration of TAP with MDFD conv achieves the best result with a PSDS1 score of 0.459, validating the complementary strengths of temporal attention and multi-scale frequency adaptation. These findings establish TFD conv as a powerful and generalizable framework for enhancing both transient sensitivity and overall feature robustness in SED.
\end{abstract}

\begin{IEEEkeywords}
Sound Event Detection, Temporal Attention Pooling, Frequency Dynamic Convolution, Time-Frequency Adaptive Feature Extraction, Attention-Based Acoustic Modeling
\end{IEEEkeywords}

\section{Introduction}
Sound event detection (SED) is a fundamental task in auditory intelligence, enabling key applications such as AI-driven perception, smart environments, and bioacoustic monitoring \cite{CASSE, DCASEtask4, crnn, sedmetrics, PSDS, freqdepinternoise, jitter}. Beyond SED, extensive research has been conducted across a broad range of auditory tasks, including speech and speaker recognition \cite{specaug, conformer, mpc, wav2vec2.0, hubert, ASP, SAP, tdyaccess, freqse, c2datt}, sound event recognition \cite{PANN, coughcam, etri, AST, beats}, and sound event localization and detection \cite{seld2019, starss22, 2022t3report}. Emerging areas such as automated audio captioning \cite{dcaseaac, clotho, chatgptaugaac}, few-shot bioacoustic detection \cite{dcasebed2024, bioacousticstrf}, and computational modeling of human auditory perception \cite{prtfnet, brainstem} further illustrate the expanding scope of auditory intelligence. In parallel, recent advances in generative modeling of sound \cite{audioldm, audiogen, vifs} have explored synthesis of sound events from label or text, offering new perspectives in sound representation learning and multimodal integration.

\begin{figure}[t]
    \centering
    \includegraphics[width=1\linewidth]{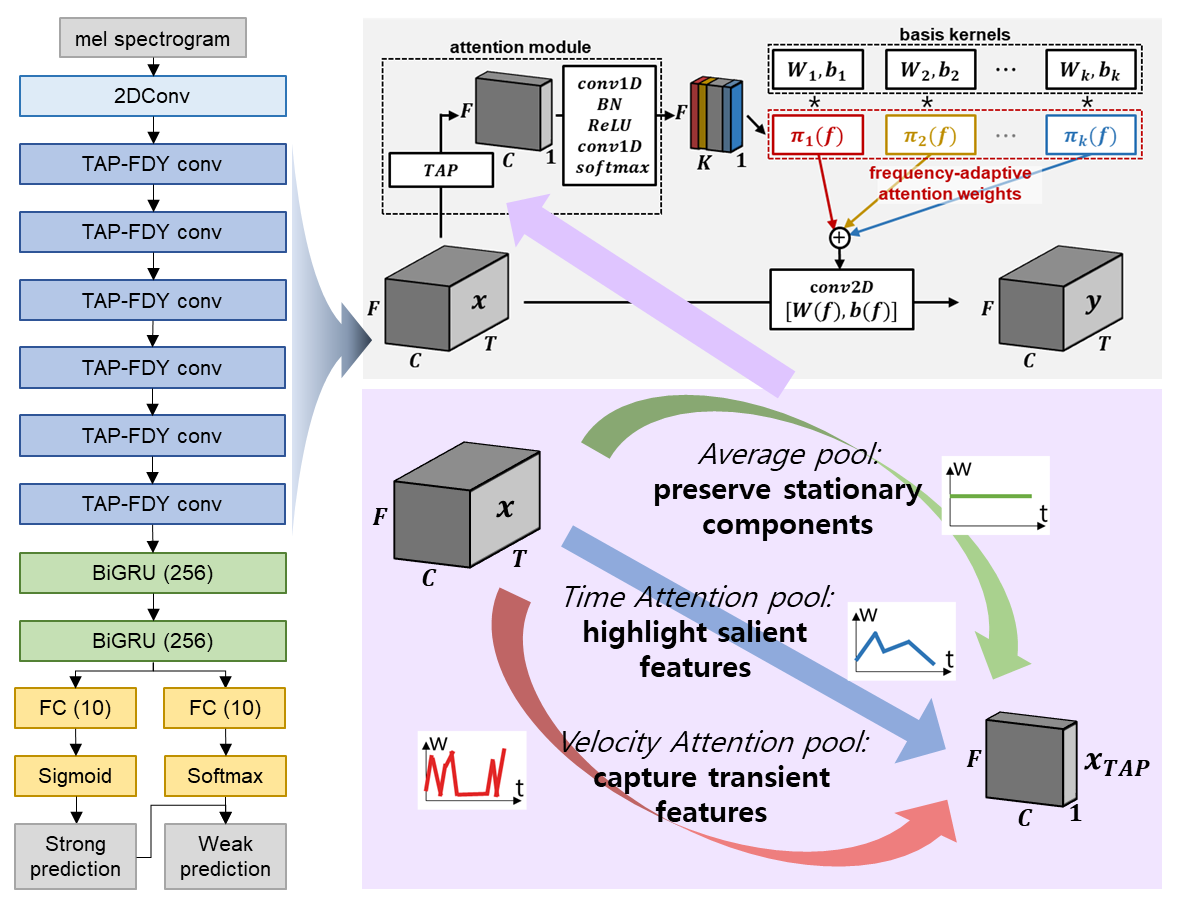}
    \caption{
        Overview of the proposed \textit{temporal attention pooling frequency dynamic convolution} (TFD conv). 
        The left side illustrates the overall architecture of the TFD conv-based SED model, where TFD conv layers replace standard FDY conv layers for enhanced time-frequency adaptive feature extraction. 
        The right side provides a detailed breakdown of the \textit{temporal attention pooling} (TAP) mechanism, which replaces temporal average pooling in FDY conv.
        TAP consists of three pooling components: 
        (a) \textit{time attention pooling} (TA), which dynamically weights salient temporal regions, 
        (b) \textit{velocity attention pooling} (VA), which applies attention based on temporal differences to emphasize transient events, and 
        (c) \textit{average pooling} to maintain robustness for stationary sound events. 
        By integrating TAP with frequency-adaptive convolution kernels, TFD conv improves the recognition of transient and quasi-stationary sound events.
    }
    \label{fig:TAP_FDY_overview}
\end{figure}

\IEEEpubidadjcol 

SED aims to identify and localize sound events within an audio signal, specifying their onset and offset times \cite{CASSE, DCASEtask4, crnn, sedmetrics, PSDS}. SED plays a crucial role in numerous real-world applications, including automated surveillance, human-computer interaction, and multimedia indexing. Additionally, studies inspired by auditory cognition have contributed to further advancements in the field \cite{filtaug, FDY, freqdeptalsp, stftchallenge, stftworkshop}. Early approaches relied on convolutional neural networks (CNNs) to capture spectral and temporal patterns in audio signals \cite{freqatt, FDY, DFD, PFD}. Recent transformer-based models also incorporate CNN branches to enhance performance \cite{dcase2023a_1st, dcase2023a_2nd, dcase2024_1st, dcase2024myworkshop, matsed, jitter, PMAM}. With recent advancements in deep learning, SED performance has significantly improved, particularly through the introduction of \textit{frequency dynamic convolution (FDY conv)} \cite{FDY, freqdeptalsp}. FDY conv enhances traditional convolution-based methods by employing frequency-adaptive kernels, enabling more precise recognition of frequency-dependent sound patterns. However, despite its state-of-the-art performance, FDY conv relies on \textit{temporal average pooling} for feature aggregation, which may hinder its ability to effectively capture transient sound events.

Temporal average pooling is widely used in FDY conv and its variants to aggregate features along the time axis by treating all temporal frames equally \cite{FDY, DFD, PFD}. Although computationally efficient and effective for preserving global temporal structure, this approach assumes that all temporal regions contribute equally to the representation of a sound event. However, many sound events exhibit transient characteristics—short, sharp bursts of sound that are crucial for detecting non-stationary events. Examples include alarm bells, door knocks, and plosive speech syllables, all of which contain essential information concentrated within narrow time ranges. Temporal average pooling tends to dilute the significance of such transient signals, resulting in suboptimal detection accuracy for these events. While FDY conv and its extensions have shown strong performance for non-stationary sound events through frequency-adaptive modeling, the uniform weighting scheme of average pooling can still hinder the optimal capture of transient temporal cues \cite{FDY}.

To address this limitation, we propose \textit{temporal attention pooling (TAP)} that adaptively weights temporal features based on their significance within the given audio context. TAP integrates three key components: \textit{time attention pooling} (TA), which dynamically assigns attention weights to salient temporal regions; \textit{velocity attention pooling} (VA), which captures transient characteristics by applying attention weights derived from temporal differences of input features; and conventional \textit{average pooling}, which ensures robustness for stationary sound events. By incorporating these mechanisms, TAP improves the balance between transient and quasi-stationary signal representation in FDY conv as shown in Fig. \ref{fig:TAP_FDY_overview}.

This paper introduces \textit{TFD conv}, which integrates TAP into FDY conv to improve time-frequency adaptive feature extraction. TFD conv replaces the average pooling layer in FDY conv with TAP, allowing for more flexible and adaptive temporal aggregation. TAP utilizes 2D convolution layers to enhance the saliency of input features before computing attention weights, ensuring more effective frequency-adaptive feature extraction. Unlike conventional pooling methods that assume uniform importance across time, TAP enables FDY conv to better capture transient sound patterns while maintaining stability for stationary signals. Furthermore, we extend our analysis by integrating TAP with previously developed FDY conv variants, including \textit{dilated FDY conv (DFD conv)}, \textit{partial FDY conv (PFD conv)}, and \textit{multi-dilated FDY conv (MDFD conv)}, to achieve a high-performing SED system on the DESED dataset without external data \cite{DFD, PFD}. Through extensive ablation studies, we evaluate the compatibility of TAP with dilated basis kernels and partial dynamic branches, demonstrating its effectiveness in enhancing both non-stationary and quasi-stationary sound event detection. Additionally, classwise performance analysis with ANOVA and Tukey HSD post-hoc testing confirms that TFD conv significantly improves detection for transient-heavy sound events compared to the other FDY-based models, which struggle with these classes due to their reliance on temporal average pooling.  

The key contributions of this work are as follows:  
\begin{enumerate}
    \itemsep 0.1em
    \item We propose \textit{TFD conv}, which replaces temporal average pooling in FDY conv with \textit{temporal attention pooling (TAP)} to improve recognition of transient sounds.
    \item TAP integrates \textit{time and velocity attention pooling} to emphasize transient cues, while retaining \textit{average pooling} to maintain robustness for stationary events.
    \item Classwise analysis using ANOVA and Tukey HSD post-hoc testing demonstrates the robustness of TFD conv across both non-stationary and quasi-stationary sounds.
    \item Extensive ablation studies confirm the compatibility of TAP with existing FDY conv variants, including DFD, PFD, and MDFD conv.
    \item Both \textit{TFD conv} and \textit{TAP + MDFD conv} achieve state-of-the-art results on the DESED dataset without external dataset or pretrained model.
\end{enumerate}
The official implementation code is available on GitHub\footnote{https://github.com/frednam93/TAP-FDY-SED}.

\section{Previous Works}
\begin{figure*}[ht]
    \centering
    \includegraphics[width=1\linewidth]{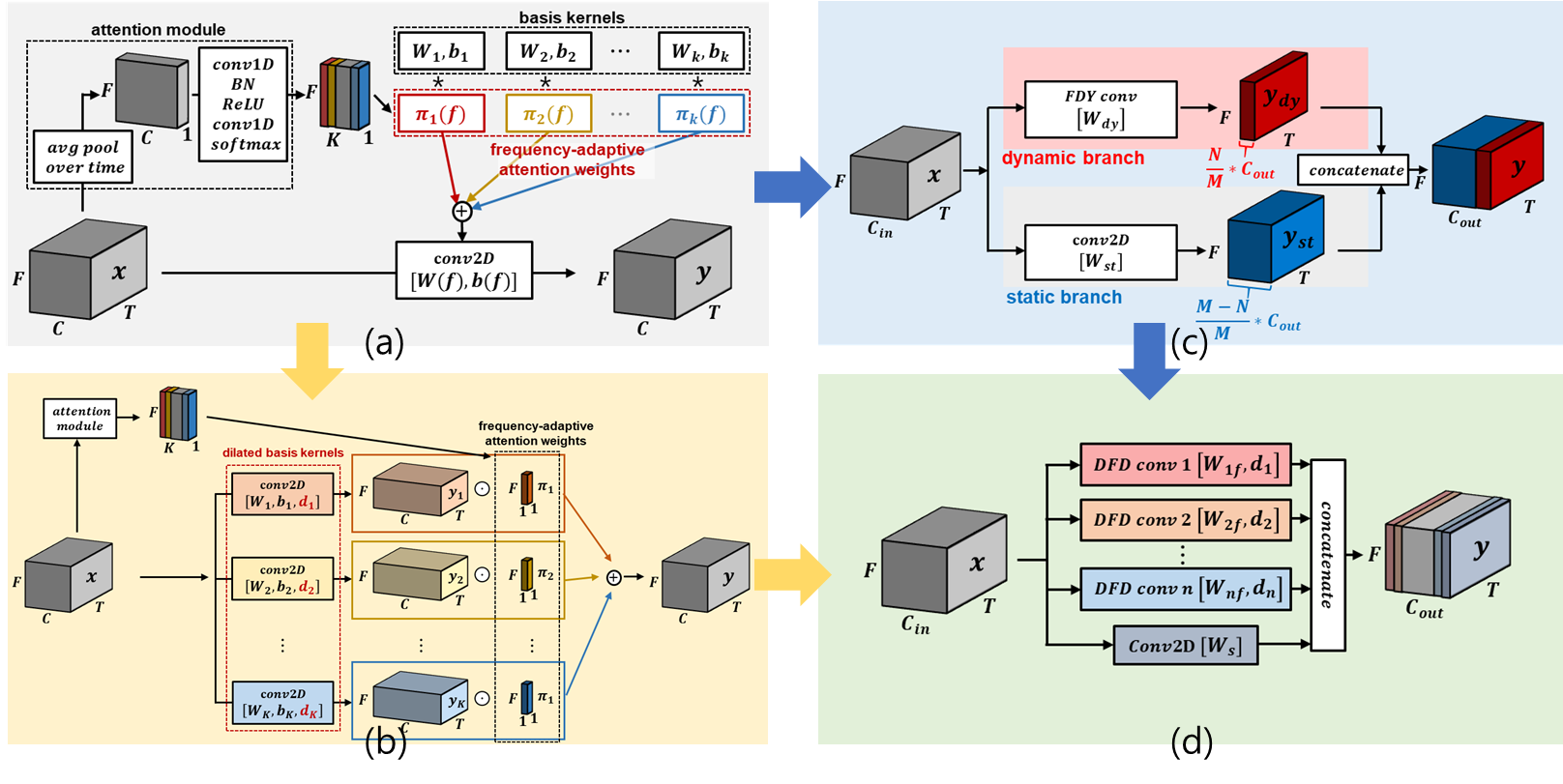}
    \caption{
        Illustration of different frequency dynamic convolution (FDY conv) variants. 
        (a) \textit{FDY conv}: Introduces frequency-adaptive convolution kernels to release the translational equivariance of conventional 2D convolution~\cite{FDY}.
        (b) \textit{DFD conv}: Incorporates dilated basis kernels to expand the spectral receptive field and diversify frequency-adaptive kernels~\cite{DFD}.
        (c) \textit{PFD conv}: Introduces a static branch alongside the FDY conv dynamic branch to reduce model complexity~\cite{PFD}.
        (d) \textit{MDFD conv}: Extends DFD and PFD by integrating multiple dilated dynamic branches within a single static branch for improved feature extraction~\cite{PFD}.
    }
    \label{fig:FDY_variants}
\end{figure*}

SED has significantly advanced with deep learning, with convolutional and recurrent neural network (CRNN) architectures remaining widely used~\cite{dcase2020_1st, dcase2021_1st, mytechreport, dcase2023a_1st, dcase2024_1st, dcase2024myworkshop}. Among them, \textit{frequency dynamic convolution (FDY conv)} has been a key development, introducing frequency-adaptive convolution kernels to release the translational equivariance constraint of conventional 2D convolution on time-frequency audio data~\cite{FDY}. This approach has emphasized the importance of frequency-dependent modeling in SED~\cite{mytechreport, filtaug, dcase2022mytechrep, freqatt}, leading to various FDY conv variants~\cite{DFD, mdfdy, PFD}. Fig.~\ref{fig:FDY_variants} illustrates the evolution of FDY conv and its extensions.

\subsection{Frequency Dynamic Convolution (FDY conv)}

FDY conv was introduced to address the inherent limitations of traditional 2D convolution when applied to time-frequency audio data~\cite{FDY}. While conventional 2D convolution assumes translation equivariance across both time and frequency axes, sound events are inherently shift-variant in the frequency domain. FDY conv resolves this issue by applying frequency-adaptive kernels, allowing the model to capture more relevant frequency-dependent patterns.

FDY conv has achieved state-of-the-art results on the DESED dataset and has been particularly effective in detecting non-stationary sound events such as speech and alarms~\cite{FDY}. It has also been widely adopted by top-performing models in the Detection and Classification of Acoustic Scenes and Events (DCASE) challenge~\cite{dcase2023a_1st, dcase2023b_1st, dcase2024mytechrep}. The overall structure of FDY conv is illustrated in Fig.~\ref{fig:FDY_variants}(a).

\subsection{Dilated Frequency Dynamic Convolution (DFD conv)}

However, FDY conv applies structurally same basis kernels, limiting the differentiation of kernel roles. Additionally, its standard 3×3 basis kernels are constrained in capturing broad frequency patterns. To further improve FDY conv, \textit{dilated frequency dynamic convolution (DFD conv)} was introduced, incorporating dilated kernels to expand the spectral receptive field and diversify frequency-adaptive kernels~\cite{DFD}. By applying varying dilation sizes to the basis kernels, DFD conv enables broader spectro-temporal feature extraction. Experimental results demonstrated that DFD-CRNN, employing dilation sizes of 1, 2, 3, and 3, outperformed FDY-CRNN, achieving a 3.12\% improvement in polyphonic sound detection score (PSDS). The structure of DFD conv is shown in Fig.~\ref{fig:FDY_variants}(b).

\subsection{Partial Frequency Dynamic Convolution (PFD conv)}

While DFD conv improved frequency modeling, FDY conv remained computationally expensive due to the increased number of basis kernels needed to form a single frequency-adaptive kernel. To address this, \textit{partial frequency dynamic convolution (PFD conv)} was proposed, introducing a static 2D convolution branch alongside the dynamic FDY conv branch~\cite{PFD}. The static branch processes conventional 2D convolution, while the dynamic branch applies frequency-adaptive kernels. The outputs of both branches are concatenated along the channel dimension, reducing model complexity while maintaining performance. Notably, setting the dynamic branch proportion to one-eighth resulted in a 51.9\% reduction in FDY-CRNN parameters without compromising performance. Fig.~\ref{fig:FDY_variants}(c) presents the structure of PFD conv.

\subsection{Multi-Dilated Frequency Dynamic Convolution (MDFD conv)}

Building upon DFD conv and PFD conv, \textit{multi-dilated frequency dynamic convolution (MDFD conv)} further enhanced feature extraction by integrating multiple dilated and non-dilated dynamic branches within a single static branch composed of a 2D convolution layer~\cite{PFD}. By leveraging multiple dynamic branches with different dilation sizes, MDFD conv improved model robustness and recognition performance. MDFD conv achieved state-of-the-art results on DESED, without using external datasets or pre-trained models~\cite{PSDS}. The overall architecture of MDFD conv is illustrated in Fig.~\ref{fig:FDY_variants}(d).

\section{Proposed Methods}
This section presents the proposed \textit{temporal attention pooling frequency dynamic convolution (TFD conv)}, which integrates \textit{temporal attention pooling (TAP)} into the FDY conv framework to address the limitations of temporal average pooling. TFD conv aims to optimally weight transient and stationary temporal features, thereby improving time-frequency adaptive feature extraction.

\subsection{Limitations of Temporal Average Pooling}

Temporal average pooling is widely used in FDY conv for feature aggregation along the time axis. While computationally efficient, it assigns equal importance to all temporal frames, potentially failing to capture transient sound events such as alarm bells, door knocks, or speech plosives. These short-duration events contain critical information concentrated within narrow temporal ranges, which may be diluted when averaged over time. Consequently, while FDY conv effectively models non-stationary sound events, its reliance on temporal average pooling may be suboptimal for capturing rapid temporal variations.

\begin{figure}[t]
    \centering
    \includegraphics[width=1\linewidth]{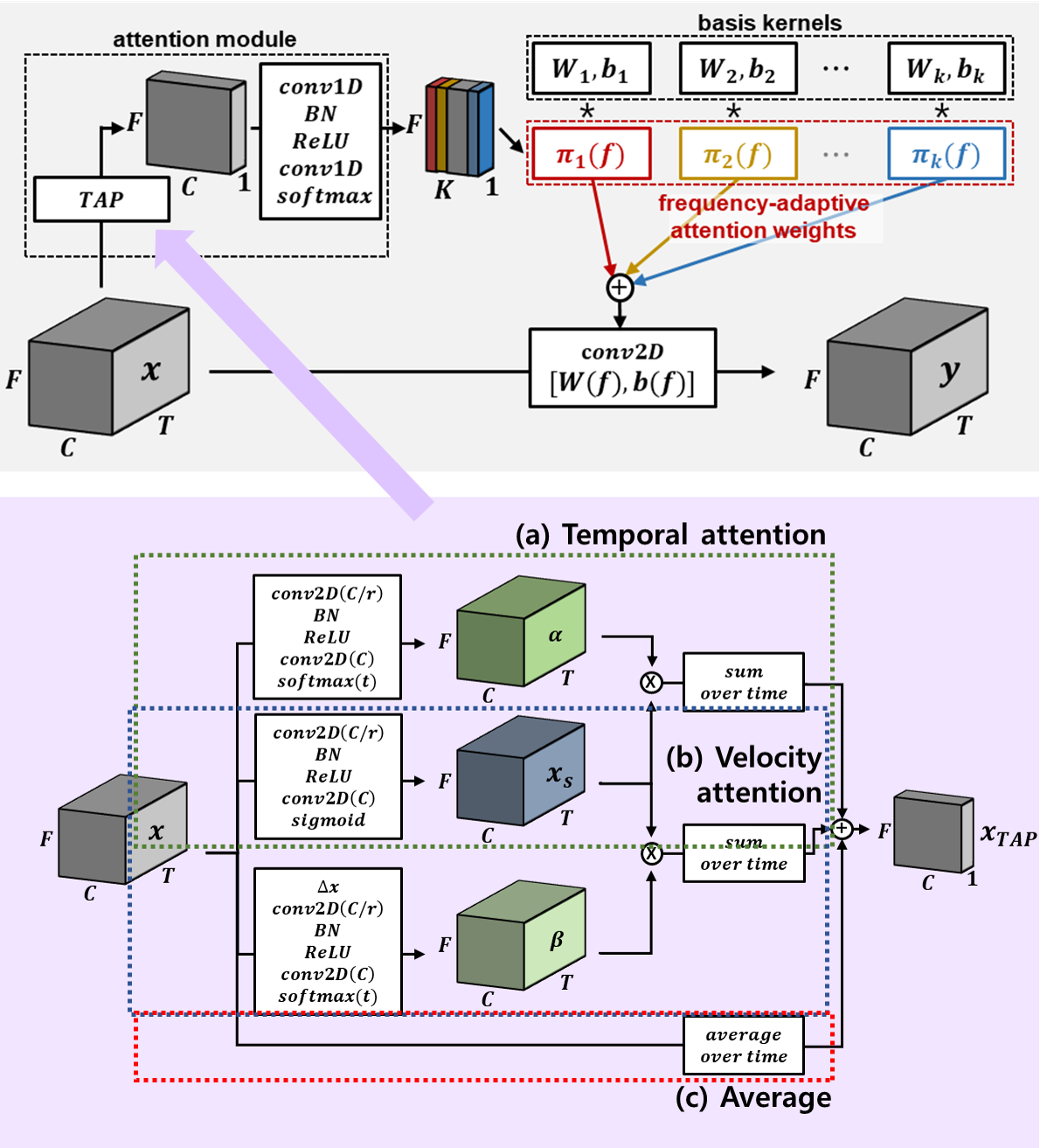}
    \caption{
        Overview of the proposed \textit{temporal attention pooling (TAP)} mechanism. TAP consists of three pooling branches: 
        (a) Attention pooling applies softmax-based attention to highlight salient temporal features.  
        (b) Velocity attention pooling incorporates temporal differences (\(\Delta x\)) and applies softmax weighting to emphasize transient sound patterns.  
        (c) Average pooling captures global temporal context by computing the mean over time.  
        The outputs of these three pooling operations are summed to obtain the final TAP feature.
    }
    \label{fig:TAPFDY}
\end{figure}

To address this limitation, we propose TAP as a replacement for average pooling in FDY conv. TAP aims to optimally weight temporal features by adaptively adjusting the contributions of transient and stationary signals. This is achieved by integrating three key pooling components: time attention pooling (TA), velocity attention pooling (VA), and average pooling. Fig.~\ref{fig:TAPFDY} illustrates the overall architecture of the proposed TAP mechanism.

\subsection{Temporal Attention Pooling (TAP)}

TAP enhances the temporal pooling mechanism by adaptively weighting features based on their significance. In this work, attention weights are computed across the time, frequency, and channel dimensions rather than being restricted to the time axis alone. This ensures that TAP captures meaningful variations across all spectro-temporal features.

The input feature tensor is denoted as:
\begin{equation}
    x \in \mathbb{R}^{C \times F \times T}
\end{equation}
where \(C\), \(F\), and \(T\) represent the number of channels, frequency bins, and temporal frames, respectively.

To enhance the extracted features, we first apply two 2D convolution operations to obtain a more salient representation:
\begin{equation}
    x_s = \text{sigmoid}(W_{s2}*\text{ReLU}(\text{BN}(W_{s1} * x + b_{s1})) + b_{s2})
\end{equation}
where \(W_{s1}\), \(b_{s1}\), \(W_{s2}\) and \(b_{s2}\) are learnable parameters, and \( * \) denotes a 2D convolution operation.

The final TAP-pooled feature is computed as:
\begin{equation}
    x_{\text{TAP}} = \sum_{t=1}^{T} \alpha_t \odot x_{s, t} 
    + \sum_{t=1}^{T} \beta_t \odot x_{s, t} 
    + \frac{1}{T} \sum_{t=1}^{T} x_t
\end{equation}
where \(x_{s, t} \in \mathbb{R}^{C \times F}\) represents the enhanced input features, and \(\alpha_t, \beta_t \in \mathbb{R}^{C \times F}\) are the attention weights from attention pooling and velocity attention pooling, respectively.

\subsubsection{Time Attention Pooling (TA)}

Attention pooling assigns weights to each time-frequency bin to modulate feature importance. As illustrated in Fig.~\ref{fig:TAPFDY}(a), attention weights \(\alpha_t\) are computed as:

\begin{equation}
    \alpha = \text{softmax}(W_{ta2}*\text{ReLU}(\text{BN}(W_{ta1} * x + b_{ta1})) + b_{ta2})
\end{equation}
where \(W_{ta1}\), \(b_{ta1}\), \(W_{ta2}\) and \(b_{ta2}\) are learnable parameters.

The time attention-pooled feature is then computed as:

\begin{equation}
    x_{\text{ta}} = \sum_{t=1}^{T} \alpha_t \odot x_{s, t}
\end{equation}

\subsubsection{Velocity Attention Pooling (VA)}

Velocity attention pooling follows a similar process to attention pooling but computes attention weights \(\beta_t\) based on temporal differences. This process is depicted in Fig.~\ref{fig:TAPFDY}(b).

\begin{equation}
    \Delta x_t = x_t - x_{t-1}
\end{equation}

The attention weights for velocity attention pooling are derived as:

\begin{equation}
    \beta = \text{softmax}(W_{va2}*\text{ReLU}(\text{BN}(W_{va1} * x + b_{va1})) + b_{va2})
\end{equation}
where \(W_{va1}\), \(b_{va1}\), \(W_{va2}\) and \(b_{va2}\) are learnable parameters.

The velocity-weighted pooled feature is then computed as:

\begin{equation}
    x_{\text{va}} = \sum_{t=1}^{T} \beta_t \odot x_{s, t}
\end{equation}

\subsubsection{Average Pooling}

To maintain robustness for stationary sound events, we retain temporal average pooling as a complementary component. The process of computing the average-pooled feature is shown in Fig.~\ref{fig:TAPFDY}(c). Notably, average pooling does not utilize $x_s$, as its purpose is to represent the time-averaged raw input feature $x$, without additional saliency enhancement. This allows the model to preserve stable temporal context useful for quasi-stationary events.

\begin{equation}
    x_{\text{avg}} = \frac{1}{T} \sum_{t=1}^{T} x_t
\end{equation}

\subsection{Integration with FDY conv Framework}

In TFD conv, the TAP feature \(x_{\text{TAP}}\) replaces the output of temporal average pooling in the FDY conv framework. This modification allows frequency-adaptive kernels to be dynamically influenced by optimally weighted temporal features. Furthermore, TAP is fully compatible with advanced FDY conv variants such as DFD conv, PFD conv, and MDFD conv, allowing seamless integration for further performance improvements.

\section{Experimental Setups}
In this section, we provide details of the SED model training framework used in this work. The overall framework follows the Detection and Classification of Acoustic Scenes and Events (DCASE) Challenge 2022 Task 4 baseline~\cite{DCASEtask4}.

\subsection{Implementation Details}

The dataset used in this work is the \textit{Domestic Environment Sound Event Detection} (DESED) dataset~\cite{DCASEtask4}, which consists of 10-second-long audio recordings sampled at 16 kHz. DESED includes real and synthetic soundscapes that simulate common domestic acoustic environments. The dataset contains ten sound event classes, including alarms, speech, running water, and vacuum cleaners. DESED consists of strongly labeled synthetic data, weakly labeled real data, and an unlabeled real dataset. The strongly labeled data contains event onset and offset annotations, whereas the weakly labeled data only indicates the presence of sound events in each clip. The unlabeled dataset has no annotations and is utilized in a semi-supervised learning framework. For training, mini-batches are constructed with 12 samples from the strongly labeled dataset, 12 from the weakly labeled dataset, and 24 from the unlabeled dataset, forming a total batch size of 48. Validation is performed using a batch of 24 samples, drawn from both the synthetic strongly labeled validation set and the weakly labeled validation set. The final evaluation uses the real validation dataset with a batch size of 24.

Semi-supervised learning is implemented using the mean teacher method~\cite{meanteacher, DCASEtask4}. Different augmentations are applied to the student and teacher models to enhance consistency learning. Training is performed for 200 epochs using the Adam optimizer, with a single NVIDIA RTX Titan GPU, except TAP+MDFD convs which used single NVIDIA A6000 GPU.

\subsection{Input Feature}

All audio waveforms are normalized such that their maximum absolute value is set to one. The log-mel spectrograms are then extracted using a short-time Fourier transform (STFT) with an FFT size of 2048, a hop length of 256, and a Hamming window. A mel filterbank with 128 mel bins is applied to the STFT magnitude output, forming the final spectrogram representation. Although the mel spectrogram's vertical axis corresponds to mel frequency bins, we refer to it as the frequency dimension throughout this paper for consistency with frequency-adaptive convolution concepts.

\subsection{Data Augmentation}

We apply multiple data augmentation techniques, including frameshift\cite{DCASEtask4}, mixup\cite{mixup}, and time masking\cite{specaug}. Mixup is applied to both strongly and weakly labeled datasets, while time masking is applied simultaneously to the input spectrogram and its corresponding label to ensure label consistency. Furthermore, FilterAugment\cite{filtaug} is employed with different filtering parameters for the student and teacher models to enhance representation diversity within the mean teacher framework.

\subsection{Baseline Model Architecture}

The baseline SED model follows the CRNN-based architecture with an attention module. The network consists of:
\begin{itemize}
    \item A CNN backbone with seven convolutional layers.
    \item Two bidirectional gated recurrent units (biGRU).
    \item A fully connected (FC) layer with an attention mechanism.
\end{itemize}
as shown in Fig. \ref{fig:TAP_FDY_overview}.

Each convolutional layer is followed by batch normalization, ReLU activation, dropout (0.5 probability), and 2D average pooling. When frequency-adaptive convolutions are applied, the second to seventh convolutional layers are replaced accordingly \cite{FDY}. The RNN module consists of two biGRU layers followed by dropout (0.5). A final FC layer with a sigmoid activation produces the frame-wise strong predictions. The attention module pools the output over the time axis to obtain weak predictions, which indicate the presence of sound events without temporal localization.

\subsection{Loss Function}

The loss function consists of four terms: strong classification loss, weak classification loss, strong consistency loss, and weak consistency loss \cite{meanteacher}. It is formulated as:

\begin{equation}
    L = BCE(SP_s, l_s) + w_w BCE(WP_s, l_w) + w_c L_C
\end{equation}
where \( BCE(x, y) \) is the binary cross-entropy (BCE) loss, \( w_w \) and \( w_c \) are the weights for the weak classification and consistency losses, \( SP_s \) and \( WP_s \) denote the strong and weak predictions of the student model, and \( l_s \) and \( l_w \) denote the strong and weak labels. The consistency loss \( L_C \) is defined as:

\begin{equation}
    L_C = MSE(SP_T, sg(SP_s)) + MSE(WP_T, sg(WP_s))
\end{equation}
where \( MSE(x, y) \) denotes mean square error (MSE) loss, and \( sg(x) \) is the stop gradient operation. The consistency loss weight \( w_c \) is increased exponentially from zero to two during the first 50 epochs.

\subsection{Post Processing}

After obtaining predictions, weak prediction masking is applied to the strong predictions, retaining strong outputs only when their confidence exceeds the corresponding weak prediction values~\cite{mytechreport}. This step ensures consistency between strong and weak predictions by filtering out low-confidence frames. Subsequently, a median filter of length 7 (approximately 0.45 seconds) is applied, following the DCASE Task 4 baseline~\cite{DCASEtask4}. Although class-wise median filtering could further optimize results, we adopt a fixed-length median filter across all classes to ensure fair comparison among models, minimizing the influence of post-processing.

\subsection{Evaluation Metrics}

To evaluate SED performance, we adopt the polyphonic sound detection score (PSDS)~\cite{PSDS}, a metric specifically designed to assess polyphonic SED systems. PSDS addresses limitations of traditional collar-based event F-scores and event error rates by incorporating intersection-based matching criteria \cite{sedmetrics}. Moreover, PSDS considers the full polyphonic receiver operating characteristic (ROC) curve, providing a robust summary of system performance across various operating points. This makes PSDS less sensitive to annotation noise and more appropriate for real-world deployment, offering better insight into classification stability across sound classes and dataset biases.

The DCASE Challenge 2021–2023 Task 4~\cite{DCASEtask4} adopts two PSDS variants: PSDS1, which focuses on accurate temporal localization, and PSDS2, which emphasizes presence/absence classification, making it more suitable for audio tagging~\cite{mytechreport, SEBB}. Since our goal is precise detection of sound event boundaries, we report only PSDS1 in this work. All PSDS1 scores in the tables correspond to the average score among twelve independent training runs, ensuring reliable and well-optimized performance evaluation of each model configuration.

\section{Results and Discussion}
\label{sec:results}
This section presents the experimental results and analysis for the proposed TFD conv. First, we evaluate the contribution of each TAP component through an ablation study. Then, we compare TFD conv with previous FDY conv variants. Finally, we investigate the effect of integrating TAP with DFD conv, PFD conv, and MDFD conv.

\subsection{Effect of Salient Representation $x_s$}

Before analyzing the full integration of temporal attention pooling (TAP), we first examine the effect of the salient representation $x_s$, which is used to enhance the spectral discriminability of features prior to attention computation. In TAP, both time attention and velocity attention branches use $x_s$ as input to compute attention weights, where $x_s$ is obtained by applying 2D convolutions to the raw feature $x$. This operation is intended to emphasize spectro-temporally meaningful features and enable the extraction of more frequency-adaptive attention weights.

To evaluate the necessity and effectiveness of this component, we compare the performance of FDY conv using attention pooling with and without $x_s$, as shown in Table~\ref{tab:x_salient}. When attention weights are computed directly from the raw feature $x$, a moderate improvement over the baseline FDY conv is observed, with PSDS1 of 0.434. However, replacing $x$ with $x_s$ further improves performance to 0.439, confirming the benefit of explicitly modeling salient representations for transient sound event detection. Although using $x$ is computationally lighter, incorporating $x_s$ offers a favorable trade-off between performance and complexity, especially when precise attention guidance is essential. All subsequent experiments adopt $x_s$ as the input to both TA and VA components.

\begin{table}[t]
    \centering
    \caption{Ablation study on salient representation $x_s$.}
    \begin{tabular}{l|c}
        \hline
        \textbf{Model} & \textbf{PSDS1} \\
        \hline
        FDY & 0.431 \\
        FDY w/ TA w/ $x$ & 0.434 \\
        FDY w/ TA w/ $x_s$ & 0.439 \\
        \hline
    \end{tabular}
    \label{tab:x_salient}
\end{table}

\subsection{Ablation Study on TAP Components}

\begin{table}[t]
    \centering
    \caption{Ablation study on \textit{temporal attention pooling (TAP)} components.}
    \begin{tabular}{l|ccc|c}
        \hline
        \textbf{Model} & \textbf{Avg} & \textbf{TA} & \textbf{VA} & \textbf{PSDS1} \\
        \hline
        FDY & \checkmark &  &  & 0.431 \\
        FDY w/ TA & & \checkmark &  & 0.439 \\
        FDY w/ VA & &  & \checkmark & 0.440 \\
        FDY w/ avg+TA & \checkmark & \checkmark &  & 0.436 \\
        FDY w/ avg+VA & \checkmark &  & \checkmark & 0.434 \\
        FDY w/ TA+VA &  & \checkmark & \checkmark & 0.440 \\
        FDY w/ avg+TA+VA (TAP, best) & \checkmark & \checkmark & \checkmark & \textbf{0.444} \\
        \hline
    \end{tabular}
    \label{tab:tap_ablation}
\end{table}

To analyze the effectiveness of each TAP component, we perform an ablation study by incrementally replacing or adding time attention pooling (TA) and velocity attention pooling (VA) to the FDY conv framework. The results are presented in Table~\ref{tab:tap_ablation}, demonstrating the importance of each component in TAP. Notably, both TA and VA contribute positively to performance, with VA alone (0.440) slightly outperforming TA alone (0.439). This suggests that capturing transient variations is particularly beneficial for sound event detection. Interestingly, the combination of TA and VA without average pooling (0.440) does not exceed the performance of VA alone, indicating thaft while TA is useful, its full potential is realized when combined with average pooling.

The best performance (0.444) is achieved when all three pooling mechanisms—TA, VA, and average pooling—are used together. This confirms that TAP effectively balances transient and stationary signal representation. The inclusion of average pooling ensures robustness to stationary signals, while TA and VA improve transient event detection.

\subsection{Comparison with FDY Conv Variants}

\begin{table}[t]
    \centering
    \caption{Comparison of different FDY convolution variants.}
    \begin{tabular}{l|c|c}
        \hline
        \textbf{Model} & \textbf{Params (M)} & \textbf{PSDS1} \\
        \hline
        Baseline (CRNN)  & 4.428 & 0.395 \\
        FDY conv  & 11.061  & 0.431 \\
        DFD conv  & 11.061  & 0.437 \\
        PFD conv  & 5.041  & 0.424 \\
        MDFD conv  & 18.157  & \textbf{0.444} \\
        TFD conv (proposed)  & 12.703  & \textbf{0.444} \\
        \hline
    \end{tabular}
    \label{tab:fdy_comparison}
\end{table}

Table~\ref{tab:fdy_comparison} compares TFD conv with previous FDY conv variants. It highlights that TFD conv achieves the highest PSDS1 score (0.444), matching MDFD conv while using significantly fewer parameters (12.703M vs. 18.157M). This suggests that TAP effectively enhances FDY conv’s feature extraction without requiring the complex multi-branch structures of MDFD conv. Compared to FDY conv, TFD conv improves PSDS1 by 3.02\%, demonstrating that TAP improves temporal feature modeling while maintaining a relatively low computational cost.

\subsection{Classwise Performance Analysis and ANOVA}  

\begin{figure}[t]
    \centering
    \includegraphics[width=1\linewidth]{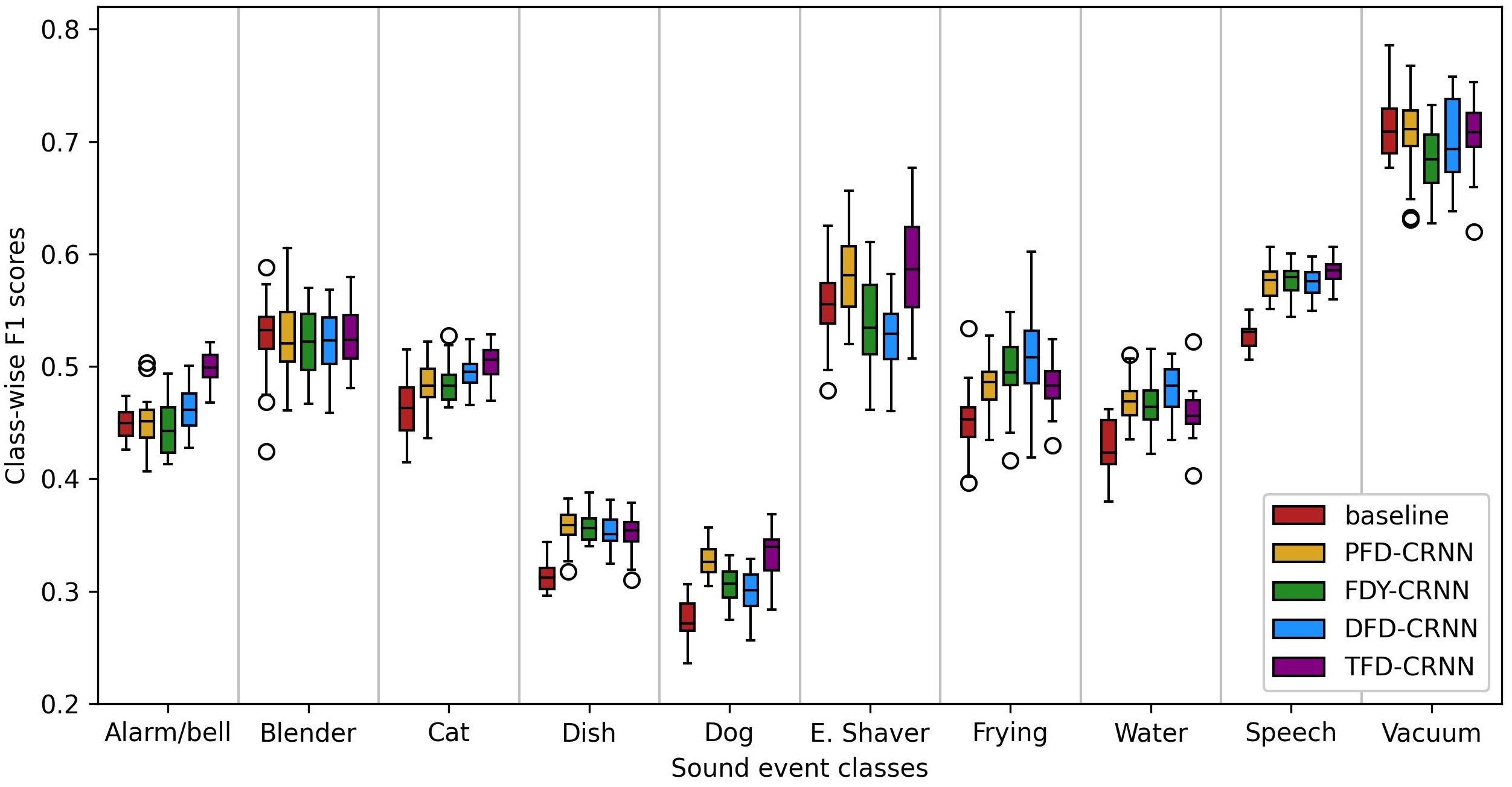}
    \caption{Classwise F1 score distribution of different models across ten sound event classes.}
    \label{fig:classwise_performance}
\end{figure}

To further investigate the effect of integrating \textit{TFD conv} on different sound event categories, we conduct a class-wise performance analysis using F1 scores. In addition, we apply ANOVA followed by Tukey’s HSD post-hoc analysis to determine statistical significance between different models. Fig. ~\ref{fig:classwise_performance} illustrates the class-wise F1 score distribution, while Table~\ref{tab:anova_results} presents the statistical comparisons.

\begin{table}[t]
    \centering
    \caption{ANOVA + Tukey HSD post-hoc analysis results for class-wise F1 scores.}
    \begin{tabular}{l|c}
        \hline
        \textbf{Class} & \textbf{ANOVA + Tukey HSD post-hoc analysis} \\
        \hline
        Alarm/bell & FDY = PFD = Baseline $\leq$ DFD $<$ TFD \\
        Blender & Baseline = PFD = FDY = DFD = TFD \\
        Cat & Baseline $<$ PFD = FDY = DFD $\leq$ TFD \\
        Dish & Baseline $<$ PFD = FDY = DFD = TFD \\
        Dog & Baseline $<$ DFD $\leq$ FDY $\approx$ PFD $\leq$ TFD \\
        Electric Shaver & FDY, DFD $\leq$ Baseline $\approx$ PFD $\leq$ TFD \\
        Frying & Baseline $<$ PFD = FDY = DFD = TFD \\
        Running Water & Baseline $<$ PFD = FDY = DFD = TFD \\
        Speech & Baseline $<$ PFD = FDY = DFD = TFD \\
        Vacuum Cleaner & FDY $\leq$ PFD = DFD = TFD $\leq$ Baseline \\
        \hline
    \end{tabular}
    \label{tab:anova_results}
\end{table}

The results in Table~\ref{tab:anova_results} reveal several key insights regarding the behavior of different models across sound event classes. First, the baseline CRNN model performs best on quasi-stationary sound events, such as electric shaver and vacuum cleaner. These sounds exhibit relatively stable spectral patterns over time, making them less reliant on improved temporal modeling. Interestingly, the baseline model outperforms all other models for vacuum cleaner, suggesting that frequency-dynamic convolutions may not provide additional benefits for highly stationary sounds.

Conversely, TFD conv (TFD-CRNN) significantly enhances the detection of non-stationary sound events, including alarm/bell, cat, dog, and electric shaver. These sound events tend to have transient characteristics with rapid spectral changes, which are better captured by the proposed TAP mechanism. The ANOVA results confirm that TFD-CRNN achieves statistically significant improvements over FDY-based models for these classes.

Among the FDY-based models, DFD-CRNN shows superior performance on alarm/Bell and vacuum cleaner compared to FDY-CRNN. This indicates that expanding the spectral receptive field using dilated convolution helps capture frequency variations critical for these classes. However, no significant improvement is observed for other classes.

PFD-CRNN demonstrates a trade-off between performance and computational efficiency. While it achieves comparable results to FDY-CRNN for most classes, it shows notable improvements for electric shaver and vacuum cleaner. This suggests that integrating static convolutional branches may enhance robustness for certain quasi-stationary events, although the overall gains are limited.

Finally, TFD-CRNN consistently outperforms all other models on transient-heavy sound events. This confirms that integrating temporal attention pooling effectively mitigates the weaknesses of temporal average pooling, enabling the model to better capture critical event cues across time.

\begin{itemize}
    \itemsep 0.1em
    \item Quasi-stationary sound events: The baseline model performs best, suggesting limited benefits from FDY-based methods for these classes.  
    \item Non-stationary sound events: TFD conv significantly outperforms other models, demonstrating the importance of attention-based temporal pooling.  
    \item DFD-CRNN excels at frequency-variant sounds but does not generalize as well to other transient sounds.  
    \item PFD-CRNN provides computational efficiency but does not consistently outperform FDY-CRNN across all classes.  
\end{itemize}

These findings further validate that TAP is especially effective in detecting non-stationary sound events, where temporal attention is essential for capturing transient cues. In addition, it simultaneously preserves strong performance on quasi-stationary events, demonstrating its versatility across diverse acoustic conditions.

\subsection{TAP + DFD Conv (Varying Dilation Sizes)}
To further explore the impact of TAP, we integrate it into DFD conv, which enhances FDY conv by applying dilation to the frequency domain. Increasing the dilation size allows the model to expand its receptive field without adding extra parameters, leading to improved kernel diversity and better adaptation to frequency-dependent variations. Since TAP primarily focuses on optimizing temporal pooling, its integration with DFD conv is expected to complement the enhanced frequency modeling, resulting in a more balanced time-frequency feature extraction.

\begin{table}[t]
    \centering
    \caption{Ablation study on TAP integrated with DFD conv (various dilation sizes).}
    \begin{tabular}{l|c|c}
        \hline
        \textbf{Model} & \textbf{Dilation Sizes} & \textbf{PSDS1} \\
        \hline
        TFD conv   & (1) & 0.444 \\
        \hline
        TAP + DFD & (2) & 0.439 \\
        TAP + DFD & (3) & 0.441 \\
        TAP + DFD & (2,3) & 0.440 \\
        TAP + DFD & (2,2,3) & 0.442 \\
        TAP + DFD & (2,3,3) & 0.440\\
        \hline
    \end{tabular}
    \label{tab:tap_dfd_ablation}
\end{table}

Table~\ref{tab:tap_dfd_ablation} presents the impact of integrating TAP with various dilation configurations in DFD conv. The results indicate that none of the dilation settings improve performance over the baseline TFD conv without dilation, suggesting that the interaction between TAP and DFD conv requires careful balance. While previous studies have shown that DFD conv is more beneficial for PSDS2—emphasizing classification over temporal localization—our findings are consistent with this, as dilation fails to improve PSDS1, which reflects precise time boundary detection~\cite{DFD}.

The baseline TFD conv (dilation size = 1) achieves a PSDS1 of 0.444, whereas the introduction of dilation leads to slight performance drops, with scores ranging from 0.439 to 0.442. This performance degradation may be attributed to the trade-off between frequency modeling and temporal resolution: while DFD conv enhances spectral receptive fields, it may also smooth out localized transient features that TAP is specifically designed to preserve.

Among the tested configurations, the (2,2,3) setting exhibits the least degradation (0.442), supporting prior findings~\cite{DFD} that more diverse dilation patterns outperform less diverse ones. This configuration appears to strike a better balance by providing sufficient frequency diversity while still preserving the temporal granularity critical for transient event detection.

These findings suggest that while TAP and DFD conv can be complementary, their combination requires careful architectural tuning. Moderate and diverse dilation sizes may enhance performance, but excessive dilation undermines TAP’s temporal sensitivity. Future research could explore adaptive or input-aware dilation strategies that dynamically balance temporal and spectral modeling depending on the characteristics of the sound event.

\subsection{TAP + PFD Conv (Varying Channel Proportion)}

To further investigate the adaptability of TAP, we integrate it into PFD conv, which reduces model complexity by balancing static and dynamic convolution branches. PFD conv achieves this by introducing a static convolutional branch alongside the frequency-adaptive dynamic branch, significantly reducing the number of parameters while maintaining competitive performance. However, reducing the proportion of dynamic channels may lead to a loss of adaptive frequency modeling capability. In this context, TAP is expected to mitigate this drawback by enhancing the temporal feature representation, compensating for the reduced dynamic processing capacity.

\begin{table}[t]
    \centering
    \caption{Ablation study on TAP integrated with PFD conv (various proportions).}
    \begin{tabular}{l|cc|c}
        \hline
        \textbf{Model} & \textbf{Channel Proportion} & \textbf{Params(M)} & \textbf{PSDS1} \\
        \hline
        Baseline (CRNN)  & 0 & 4.428 & 0.395 \\
        PFD  & 1/8 & 5.041 & 0.424 \\
        TAP + PFD & 1/32 & 6.435 & 0.428 \\
        TAP + PFD & 1/16 & 6.637 & 0.422 \\
        TAP + PFD & 1/8 & 7.042 & 0.423 \\
        TAP + PFD & 2/8 & 7.850 & 0.432 \\
        TAP + PFD & 3/8 & 8.659 & 0.431 \\
        TAP + PFD & 4/8 & 9.468 & 0.434 \\
        TAP + PFD & 5/8 & 10.276 & 0.441 \\
        TAP + PFD & 6/8 & 11.085 & 0.437 \\
        TAP + PFD & 7/8 & 11.894 & 0.431 \\
        TFD conv & 8/8 & 12.703 & \textbf{0.444} \\
        \hline
    \end{tabular}
    \label{tab:tap_pfd_ablation}
\end{table}

Table~\ref{tab:tap_pfd_ablation} presents the results of integrating TAP into PFD conv while varying the proportion of dynamic channels. The performance exhibits a non-linear trend depending on the dynamic-to-static channel ratio. The best performance (PSDS1 = 0.441) is observed when 5/8 of the channels are dynamic, suggesting that an optimal balance between dynamic and static branches is essential. A lower proportion of dynamic channels (e.g., 1/32 or 1/16) leads to noticeable performance degradation, likely due to insufficient frequency adaptivity in static branches. As the dynamic ratio increases, the model’s ability to handle complex frequency variations improves, culminating in a PSDS1 of 0.444 at the 8/8 configuration—equivalent to TFD conv.

These findings indicate that TAP can successfully complement PFD conv by enhancing temporal modeling, allowing reduced-parameter models to achieve performance levels close to fully dynamic configurations. While PFD conv helps reduce model complexity, excessively static configurations are suboptimal as they fail to capture intricate spectral patterns. The results underscore the synergy between frequency-adaptive convolution and temporal attention pooling, emphasizing that both frequency and temporal adaptivity are critical to achieving robust and efficient SED systems.

\subsection{TAP + MDFD Conv (Various Configurations)}

To further evaluate the robustness of TAP, we integrate it into MDFD conv, which extends FDY conv by incorporating multiple dilated dynamic branches. MDFD conv is designed to enhance frequency-adaptive kernel diversity by employing various dilation sizes across different dynamic branches, allowing the model to capture multi-scale spectral patterns effectively. While MDFD conv significantly improves frequency modeling, its reliance on temporal average pooling may still limit its ability to properly represent transient sound events. By introducing TAP into MDFD conv, we aim to refine the model’s temporal sensitivity while preserving its strong frequency adaptability.

\begin{table*}[t]
    \centering
    \caption{Ablation study on TAP integrated with MDFD conv (various configurations).}
    \begin{tabular}{l|c|c|c|c}
        \hline
        \textbf{Model} & \textbf{\# Channels} & \textbf{Dilation Sizes} & \textbf{Params (M)} & \textbf{PSDS1} \\
        \hline
        TFD conv    & 1/1 (32, 64, 128, 256) & (1) & 12.703 & 0.444 \\
        MDFD(1/8)  & 11/8 (44, 88, 176, 352) & (1)×5+(2,3)+(2,2,3)+(2,3,3) & 18.157 & 0.444 \\
        \hline
        TAP + MDFD(1/4) & 4/4 (32, 64, 128, 256) & (1)×2 & 11.274 & 0.429 \\
        TAP + MDFD(1/4) & 4/4 (32, 64, 128, 256) & (1)×3 & 14.697 & 0.440 \\
        \hline
        TAP + MDFD(1/8) & 8/8 (32, 64, 128, 256) & (1)×5 & 17.501 & 0.439 \\
        TAP + MDFD(1/8) & 8/8 (32, 64, 128, 256) & (1)×6 & 20.116 & 0.438 \\
        \hline
        TAP + MDFD(1/16) & 16/16 (32, 64, 128, 256) & (1)×10 & 26.532 & 0.438 \\
        TAP + MDFD(1/16) & 16/16 (32, 64, 128, 256) & (1)×11 & 28.742 & 0.434 \\
        TAP + MDFD(1/16) & 16/16 (32, 64, 128, 256) & (1)×12 & 30.953 & 0.432 \\
        \hline
        \hline
        TAP + MDFD(1/4) & 5/4 (40, 80, 160, 320) & (1)×4 & 25.266 & 0.443 \\
        TAP + MDFD(1/4) & 6/4 (48, 96, 192, 384) & (1)×5 & 40.100 & 0.435 \\
        \hline
        TAP + MDFD(1/4) & 5/4 (40, 80, 160, 320) & (1)×3+(2,2,3) & 25.266 & 0.442 \\
        TAP + MDFD(1/4) & 5/4 (40, 80, 160, 320) & (1)×3+(2,3,3) & 25.266 & \textbf{0.445} \\
        TAP + MDFD(1/4) & 5/4 (40, 80, 160, 320) & (1)×2+(2,2,3)+(2,3,3) & 25.266 & 0.444 \\
        TAP + MDFD(1/4) & 5/4 (40, 80, 160, 320) & (1)+(2,3)+(2,2,3)+(2,3,3) & 25.266 & 0.439 \\
        \hline
    \end{tabular}
    \label{tab:tap_mdfd_ablation}
\end{table*}

Table~\ref{tab:tap_mdfd_ablation} presents the ablation results of TAP integrated with MDFD conv under various configurations of channel allocations and dilation sizes. Several configurations with TAP + MDFD led to noticeable performance degradation, particularly when the number of dynamic branches increased. This suggests that excessive expansion of dynamic branches may introduce redundancy or noise into the attention pooling process, ultimately hindering the model's ability to capture transient features.

Consistent with prior findings, the use of a static branch appears beneficial \cite{PFD}. For example, the optimal MDFD conv configuration without dilation employed three non-dilated dynamic branches and 1/4 channel allocation, balancing static and dynamic computation. However, none of the TAP + MDFD configurations using non-dilated dynamic branches outperformed the simpler TFD conv baseline, indicating that TAP already provides strong temporal modeling, diminishing the marginal utility of additional non-dilated paths.

Among all configurations, the best result (PSDS1 = 0.445) was achieved with TAP + MDFD using a 5/4 channel setup and a moderate dilation scheme of (1)×3+(2,2,3). This highlights that moderate dilation diversity, when paired with TAP, enables a more effective balance between frequency-adaptive modeling and transient-aware temporal pooling.

Another key observation is the impact of channel scaling. The best-performing TAP + MDFD models employed wider channel setups (e.g., 40, 80, 160, 320), suggesting that richer representational capacity in each stage helps maximize the benefit of TAP. In contrast, extremely deep or aggressively dilated configurations (e.g., 16/16 channels with (1)×12 or 5/4 channels with (1)+(2,3)+(2,2,3)+(2,3,3)) tended to suffer from optimization challenges and over-smoothing effects.

In summary, although TAP enhances temporal modeling, its integration with MDFD requires careful architectural design. A moderate number of well-dilated branches, balanced channel growth, and restrained structural complexity are essential to fully leverage the complementary strengths of TAP and MDFD conv.

\subsection{Maximum PSDS Comparison Across FDY Conv Variants}

\begin{table}[t]
    \centering
    \caption{Comparison of different FDY convolution variants with maximum PSDS1 scores.}
    \begin{tabular}{l|c|c}
        \hline
        \textbf{Model} & \textbf{Params (M)} & \textbf{Max PSDS1} \\
        \hline
        Baseline (CRNN)  & 4.428 & 0.410 \\
        FDY conv  & 11.061  & 0.441 \\
        DFD conv  & 11.061  & 0.448 \\
        PFD conv  & 5.041  & 0.441 \\
        MDFD conv  & 18.157  & 0.455 \\
        TFD conv   & 12.703  & 0.456 \\
        TAP + MDFD & 25.266 & \textbf{0.459} \\
        \hline
    \end{tabular}
    \label{tab:max_psds}
\end{table}

To further assess the effectiveness of TFD conv, we compare the maximum PSDS1 scores achieved by different FDY conv variants, as shown in Table~\ref{tab:max_psds}. Notably, the proposed TAP + MDFD configuration achieves the highest score of 0.459, surpassing all baseline and advanced FDY conv models. This result confirms that integrating \textit{temporal attention pooling} (TAP) with a well-balanced \textit{multi-dilated frequency dynamic convolution} (MDFD) structure can further enhance performance beyond either component alone.

While TFD conv alone achieves a strong PSDS1 score of 0.456 with moderate model complexity (12.703M parameters), its combination with MDFD conv pushes the boundary even further, reaching state-of-the-art performance at 0.459. These improvements are attributed to TAP's ability to adaptively emphasize both transient and stationary sound components—effectively mitigating the over-smoothing behavior of temporal average pooling—while MDFD contributes by diversifying frequency representations through multi-scale receptive fields.

Importantly, the observed performance gain in TAP + MDFD underscores that TAP generalizes well across FDY conv structures, even in highly dynamic, multi-branch configurations. This demonstrates that TAP not only enhances temporal modeling but also scales robustly when fused with architectures focused on spectral diversity.

\begin{itemize}
    \itemsep 0.1em
    \item TAP + MDFD achieves the highest PSDS1 score of 0.459 among all evaluated models.
    \item TAP adaptively weights transient and stationary features, overcoming limitations of average pooling.
    \item The synergy between TAP and MDFD conv enables better time-frequency feature extraction without excessive model complexity.
    \item TAP exhibits strong generalizability and integration flexibility across diverse FDY conv variants.
\end{itemize}

\section{Conclusion}
In this paper, we proposed the \textit{temporal attention pooling frequency dynamic convolution} (TFD conv), an enhanced variant of FDY conv that replaces temporal average pooling with a more flexible \textit{temporal attention pooling} (TAP) mechanism. Composed of \textit{time attention pooling (TA)}, \textit{velocity attention pooling (VA)}, and \textit{average pooling}, TAP enables adaptive weighting of temporal features, effectively capturing both transient and stationary components in sound events. Comprehensive ablation studies and classwise analyses demonstrate that TFD conv improves PSDS1 by 3.02\% and significantly boosts performance on transient-heavy events such as \textit{alarm/bell} and \textit{speech}, while maintaining robustness on quasi-stationary classes, where FDY conv has been failed to outperform conventional 2D convolution. We further validated the compatibility of TAP with advanced FDY conv variants—DFD, PFD, and MDFD conv. The best-performing configuration, TAP + MDFD conv, achieved a PSDS1 of 0.459, surpassing all prior FDY conv-based systems. These results confirm that TFD conv is a generalizable and effective approach for SED, providing a robust solution across diverse temporal and spectral dynamics.

\vspace{11pt}

\bibliographystyle{IEEEtran}
\bibliography{refs}

\begin{thebibliography}{10}
\providecommand{\url}[1]{#1}
\csname url@samestyle\endcsname
\providecommand{\newblock}{\relax}
\providecommand{\bibinfo}[2]{#2}
\providecommand{\BIBentrySTDinterwordspacing}{\spaceskip=0pt\relax}
\providecommand{\BIBentryALTinterwordstretchfactor}{4}
\providecommand{\BIBentryALTinterwordspacing}{\spaceskip=\fontdimen2\font plus
\BIBentryALTinterwordstretchfactor\fontdimen3\font minus \fontdimen4\font\relax}
\providecommand{\BIBforeignlanguage}[2]{{%
\expandafter\ifx\csname l@#1\endcsname\relax
\typeout{** WARNING: IEEEtran.bst: No hyphenation pattern has been}%
\typeout{** loaded for the language `#1'. Using the pattern for}%
\typeout{** the default language instead.}%
\else
\language=\csname l@#1\endcsname
\fi
#2}}
\providecommand{\BIBdecl}{\relax}
\BIBdecl

\bibitem{CASSE}
T.~Virtanen, M.~D. Plumbley, and D.~Ellis, \emph{Computational Analysis of Sound Scenes and Events}, 1st~ed.\hskip 1em plus 0.5em minus 0.4em\relax Springer Publishing Company, Incorporated, 2017, pp. 3--11, 71--77.

\bibitem{DCASEtask4}
N.~Turpault, R.~Serizel, A.~P.~Shah, and J.~Salamon, ``{Sound event detection in domestic environments with weakly labeled data and soundscape synthesis},'' in \emph{{DCASE Workshop}}, 2019.

\bibitem{crnn}
E.~{\c{C}}ak{\i}r, G.~Parascandolo, T.~Heittola, H.~Huttunen, and T.~Virtanen, ``Convolutional recurrent neural networks for polyphonic sound event detection,'' \emph{IEEE/ACM Transactions on Audio, Speech, and Language Processing}, vol.~25, no.~6, pp. 1291--1303, 2017.

\bibitem{sedmetrics}
A.~Mesaros, T.~Heittola, and T.~Virtanen, ``Metrics for polyphonic sound event detection,'' \emph{Applied Sciences}, vol.~6, no.~6, 2016.

\bibitem{PSDS}
{\c{C}}.~Bilen, G.~Ferroni, F.~Tuveri, J.~Azcarreta, and S.~Krstulovi\'{c}, ``A framework for the robust evaluation of sound event detection,'' in \emph{ICASSP}, 2020, pp. 61--65.

\bibitem{freqdepinternoise}
H.~Nam, S.-H. Kim, B.-Y. Ko, D.~Min, and Y.-H. Park, ``Study on frequency dependent convolution methods for sound event detection,'' in \emph{{Proc. INTER-NOISE}}, 2024.

\bibitem{jitter}
H.~Nam and Y.-H. Park, ``Jitter: Jigsaw temporal transformer for event reconstruction for self-supervised sound event detection,'' \emph{arXiv preprint arXiv:2502.20857}, 2025.

\bibitem{specaug}
D.~S. Park, W.~Chan, Y.~Zhang, C.-C. Chiu, B.~Zoph, E.~D. Cubuk, and Q.~V. Le, ``{SpecAugment: A Simple Data Augmentation Method for Automatic Speech Recognition},'' in \emph{Proc. Interspeech}, 2019.

\bibitem{conformer}
A.~Gulati, J.~Qin, C.-C. Chiu, N.~Parmar, Y.~Zhang, J.~Yu, W.~Han, S.~Wang, Z.~Zhang, Y.~Wu, and R.~Pang, ``{Conformer: Convolution-augmented Transformer for Speech Recognition},'' in \emph{Proc. Interspeech}, 2020.

\bibitem{mpc}
H.~Nam and Y.-H. Park, ``Coherence-based phonemic analysis on the effect of reverberation to practical automatic speech recognition,'' \emph{Applied Acoustics}, vol. 227, p. 110233, 2025.

\bibitem{wav2vec2.0}
A.~Baevski, Y.~Zhou, A.~Mohamed, and M.~Auli, ``wav2vec 2.0: A framework for self-supervised learning of speech representations,'' in \emph{Advances in Neural Information Processing Systems}, 2020.

\bibitem{hubert}
W.-N. Hsu, B.~Bolte, Y.-H.~H. Tsai, K.~Lakhotia, R.~Salakhutdinov, and A.~Mohamed, ``Hubert: Self-supervised speech representation learning by masked prediction of hidden units,'' \emph{IEEE/ACM Transactions on Audio, Speech, and Language Processing}, 2021.

\bibitem{ASP}
K.~Okabe, T.~Koshinaka, and K.~Shinoda, ``Attentive statistics pooling for deep speaker embedding,'' in \emph{Proc. Interspeech}, 2018.

\bibitem{SAP}
W.~Cai, J.~Chen, and M.~Li, ``Exploring the encoding layer and loss function in end-to-end speaker and language recognition system,'' in \emph{Proc. Interspeech}, 2018.

\bibitem{tdyaccess}
S.-H. Kim, H.~Nam, and Y.-H. Park, ``Analysis-based optimization of temporal dynamic convolutional neural network for text-independent speaker verification,'' \emph{IEEE Access}, vol.~11, 2023.

\bibitem{freqse}
J.~Thienpondt, B.~Desplanques, and K.~Demuynck, ``Integrating frequency translational invariance in tdnns and frequency positional information in 2d resnets to enhance speaker verification,'' in \emph{Proc. Interspeech}, 2021.

\bibitem{c2datt}
J.~Li, Y.~Tian, and T.~Lee, ``Convolution-based channel-frequency attention for text-independent speaker verification,'' in \emph{ICASSP}, 2023.

\bibitem{PANN}
Q.~Kong, Y.~Cao, T.~Iqbal, Y.~Wang, W.~Wang, and M.~D. Plumbley, ``Panns: Large-scale pretrained audio neural networks for audio pattern recognition,'' \emph{IEEE/ACM Transactions on Audio, Speech, and Language Processing}, 2020.

\bibitem{coughcam}
G.-T. Lee, H.~Nam, S.-H. Kim, S.-M. Choi, Y.~Kim, and Y.-H. Park, ``Deep learning based cough detection camera using enhanced features,'' \emph{Expert Systems with Applications}, vol. 206, 2022.

\bibitem{etri}
S.-H. Kim, H.~Nam, S.-M. Choi, and Y.-H. Park, ``Real-time sound recognition system for human care robot considering custom sound events,'' \emph{IEEE Access}, vol.~12, 2024.

\bibitem{AST}
Y.~Gong, Y.-A. Chung, and J.~Glass, ``Ast: Audio spectrogram transformer,'' in \emph{Proc. Interspeech}, 2021.

\bibitem{beats}
S.~Chen, Y.~Wu, C.~Wang, S.~Liu, D.~Tompkins, Z.~Chen, W.~Che, X.~Yu, and F.~Wei, ``Beats: Audio pre-training with acoustic tokenizers,'' in \emph{ICML}, 2023.

\bibitem{seld2019}
A.~Politis, A.~Mesaros, S.~Adavanne, T.~Heittola, and T.~Virtanen, ``Overview and evaluation of sound event localization and detection in dcase 2019,'' \emph{IEEE/ACM Transactions on Audio, Speech, and Language Processing}, vol.~29, 2020.

\bibitem{starss22}
A.~Politis, K.~Shimada, P.~Sudarsanam, S.~Adavanne, D.~Krause, Y.~Koyama, N.~Takahashi, S.~Takahashi, Y.~Mitsufuji, and T.~Virtanen, ``{STARSS22}: {A} dataset of spatial recordings of real scenes with spatiotemporal annotations of sound events,'' in \emph{{DCASE Workshop}}, 2022.

\bibitem{2022t3report}
B.-Y. Ko, H.~Nam, S.-H. Kim, D.~Min, S.-D. Choi, and Y.-H. Park, ``Data augmentation and squeeze-and-excitation network on multiple dimension for sound event localization and detection in real scenes,'' DCASE Challenge, Tech. Rep., 2022.

\bibitem{dcaseaac}
K.~Drossos, S.~Adavanne, and T.~Virtanen, ``Automated audio captioning with recurrent neural networks,'' in \emph{IEEE Workshop on Applications of Signal Processing to Audio and Acoustics}, 2017.

\bibitem{clotho}
K.~Drossos, S.~Lipping, and T.~Virtanen, ``Clotho: an audio captioning dataset,'' in \emph{ICASSP}, 2020.

\bibitem{chatgptaugaac}
I.~Choi, H.~Nam, D.~Min, S.-D. Choi, and Y.-H. Park, ``Chatgpt caption paraphrasing and fense-based caption filtering for automated audio captioning,'' DCASE Challenge, Tech. Rep., 2024.

\bibitem{dcasebed2024}
J.~Liang, I.~Nolasco, B.~Ghani, H.~Phan, E.~Benetos, and D.~Stowell, ``{Mind the Domain Gap: a Systematic Analysis on Bioacoustic Sound Event Detection},'' \emph{arXiv preprint arXiv:2403.18638}, 2024.

\bibitem{bioacousticstrf}
D.~Min, H.~Nam, and Y.-H. Park, ``Few-shot bioacoustic event detection utilizing spectro-temporal receptive field,'' in \emph{{Proc. INTER-NOISE}}, 2024.

\bibitem{prtfnet}
B.-Y. Ko, G.-T. Lee, H.~Nam, and Y.-H. Park, ``Prtfnet: Hrtf individualization for accurate spectral cues using a compact prtf,'' \emph{IEEE Access}, vol.~11, 2023.

\bibitem{brainstem}
B.-Y. Ko, Y.-H. Park, G.-T. Lee, and H.~Nam, ``Filteraugment: An acoustic environmental data augmentation method,'' in \emph{nternational Congress on Acoustics (ICA)}, 2022.

\bibitem{audioldm}
H.~Liu, Z.~Chen, Y.~Yuan, X.~Mei, X.~Liu, D.~Mandic, W.~Wang, and M.~D. Plumbley, ``{A}udio{LDM}: Text-to-audio generation with latent diffusion models,'' in \emph{ICML}, 2023.

\bibitem{audiogen}
F.~Kreuk, G.~Synnaeve, A.~Polyak, U.~Singer, A.~D{\'e}fossez, J.~Copet, D.~Parikh, Y.~Taigman, and Y.~Adi, ``Audiogen: Textually guided audio generation,'' in \emph{International Conference on Learning Representations (ICLR)}, 2023.

\bibitem{vifs}
J.~Lee, H.~Nam, and Y.-H. Park, ``Vifs: An end-to-end variational inference for foley sound synthesis,'' DCASE Challenge, Tech. Rep., 2023.

\bibitem{filtaug}
H.~Nam, S.-H. Kim, and Y.-H. Park, ``Filteraugment: An acoustic environmental data augmentation method,'' in \emph{ICASSP}, 2022.

\bibitem{FDY}
H.~Nam, S.-H. Kim, B.-Y. Ko, and Y.-H. Park, ``{Frequency Dynamic Convolution: Frequency-Adaptive Pattern Recognition for Sound Event Detection},'' in \emph{Proc. Interspeech}, 2022.

\bibitem{freqdeptalsp}
H.~Nam, S.-H. Kim, D.~Min, B.-Y. Ko, and Y.-H. Park, ``Towards understanding of frequency dependence on sound event detection,'' \emph{arXiv preprint arXiv:2502.07208}, 2025.

\bibitem{stftchallenge}
D.~Min, H.~Nam, and Y.-H. Park, ``Application of spectro-temporal receptive field on soft labeled sound event detection,'' DCASE Challenge, Tech. Rep., 2023.

\bibitem{stftworkshop}
------, ``Auditory neural response inspired sound event detection based on spectro-temporal receptive field,'' in \emph{{DCASE Workshop}}, 2023.

\bibitem{freqatt}
H.~Nam, S.-H. Kim, D.~Min, and Y.-H. Park, ``Frequency \& channel attention for computationally efficient sound event detection,'' in \emph{{DCASE Workshop}}, 2023.

\bibitem{DFD}
H.~Nam, S.-H. Kim, D.~Min, J.~Lee, and Y.-H. Park, ``Diversifying and expanding frequency-adaptive convolution kernels for sound event detection,'' in \emph{Proc. Interspeech}, 2024.

\bibitem{PFD}
H.~Nam and Y.-H. Park, ``Pushing the limit of sound event detection with multi-dilated frequency dynamic convolution,'' \emph{arXiv preprint arXiv:2406.13312}, 2024.

\bibitem{dcase2023a_1st}
J.~W. Kim, S.~W. Son, Y.~Song, H.~K. Kim, I.~H. Song, and J.~E. Lim, ``Semi-supervised learning-based sound event detection using frequency dynamic convolution with large kernel attention for {DCASE} challenge 2023 task 4,'' DCASE Challenge, Tech. Rep., 2023.

\bibitem{dcase2023a_2nd}
S.~Xiao, J.~Shen, A.~Hu, X.~Zhang, P.~Zhang, and Y.~Yan, ``Sound event detection with weak prediction for dcase 2023 challenge task4a,'' DCASE Challenge, Tech. Rep., 2023.

\bibitem{dcase2024_1st}
F.~Schmid, P.~Primus, T.~Morocutti, J.~Greif, and G.~Widmer, ``Improving audio spectrogram transformers for sound event detection through multi-stage training,'' DCASE2024 Challenge, Tech. Rep., 2024.

\bibitem{dcase2024myworkshop}
H.~Nam, D.~Min, I.~Choi, S.-D. Choi, and Y.-H. Park, ``Self training and ensembling frequency dependent networks with coarse prediction pooling and sound event bounding boxes,'' in \emph{{DCASE Workshop}}, 2024.

\bibitem{matsed}
P.~Cai, Y.~Song, K.~Li, H.~Song, and I.~McLoughlin, ``Mat-sed: A masked audio transformer with masked-reconstruction based pre-training for sound event detection,'' in \emph{Proc. Interspeech}, 2024.

\bibitem{PMAM}
P.~Cai, Y.~Song, N.~Jiang, Q.~Gu, and I.~McLoughlin, ``Prototype based masked audio model for self-supervised learning of sound event detection,'' \emph{arXiv preprint arXiv:2409.17656}, 2024.

\bibitem{dcase2020_1st}
K.~Miyazaki, T.~Komatsu, T.~Hayashi, S.~Watanabe, T.~Toda, and K.~Takeda, ``Convolution-augmented transformer for semi-supervised sound event detection,'' DCASE Challenge, Tech. Rep., 2020.

\bibitem{dcase2021_1st}
X.~Zheng, H.~Chen, and Y.~Song, ``Zheng ustc team’s submission for dcase2021 task4 – semi-supervised sound event detection,'' DCASE Challenge, Tech. Rep., 2021.

\bibitem{mytechreport}
H.~Nam, B.-Y. Ko, G.-T. Lee, S.-H. Kim, W.-H. Jung, S.-M. Choi, and Y.-H. Park, ``Heavily augmented sound event detection utilizing weak predictions,'' DCASE Challenge, Tech. Rep., 2021.

\bibitem{dcase2022mytechrep}
H.~Nam, S.-H. Kim, D.~Min, B.-Y. Ko, S.-D. Choi, and Y.-H. Park, ``Frequency dependent sound event detection for dcase 2022 challenge task 4,'' DCASE Challenge, Tech. Rep., 2022.

\bibitem{mdfdy}
S.~Xiao, X.~Zhang, and P.~Zhang, ``Multi-dimensional frequency dynamic convolution with confident mean teacher for sound event detection,'' in \emph{ICASSP}, 2023.

\bibitem{dcase2023b_1st}
H.~Yin, J.~Bai, S.~Huang, and J.~Chen, ``How information on soft labels and hard labels mutually benefits sound event detection tasks,'' DCASE Challenge, Tech. Rep., 2023.

\bibitem{dcase2024mytechrep}
H.~Nam, D.~Min, I.~Choi, S.-D. Choi, and Y.-H. Park, ``Self training and ensembling frequency dependent networks with coarse prediction pooling and sound event bounding boxes,'' DCASE Challenge, Tech. Rep., 2024.

\bibitem{meanteacher}
A.~Tarvainen and H.~Valpola, ``Mean teachers are better role models: Weight-averaged consistency targets improve semi-supervised deep learning results,'' in \emph{Advances in Neural Information Processing Systems}, vol.~30, 2017.

\bibitem{mixup}
H.~Zhang, M.~Cisse, Y.~N. Dauphin, and D.~Lopez-Paz, ``mixup: Beyond empirical risk minimization,'' in \emph{International Conference on Learning Representations (ICLR)}, 2018.

\bibitem{SEBB}
J.~Ebbers, F.~G. Germain, G.~Wichern, and J.~L. Roux, ``Sound event bounding boxes,'' in \emph{Proc. Interspeech}, 2024.

\end{thebibliography}

\begin{IEEEbiography}[{\includegraphics[width=1in,height=1.25in,clip,keepaspectratio]{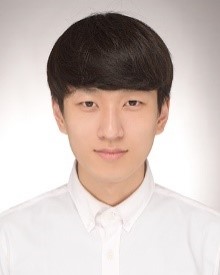}}]{Hyeonuk Nam}
received B.S. and M.S. degrees in mechanical engineering from Korea Advanced Institute of Science and Technology, Daejeon, Korea, in 2018 and 2020 respectively. He is currently pursuing the Ph.D. degree in mechanical engineering at the same institute.
His research interests include a wide range of topics in auditory intelligence, such as sound event detection, audio-language multimodal learning, self-supervised audio representation learning, audio question answering, and audio captioning.
\end{IEEEbiography}

\vspace{11pt}

\begin{IEEEbiography}[{\includegraphics[width=1in,height=1.25in,clip,keepaspectratio]{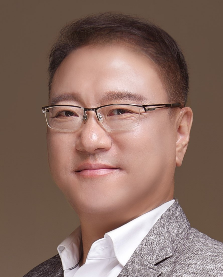}}]{Yong-Hwa Park}
received BS, MS, and PhD in Mechanical Engineering from KAIST in 1991, 1993, and 1999, respectively. In 2000, he joined to Aerospace Department at the University of Colorado at Boulder as a research associate. From 2003-2016, he worked for Samsung Electronics in the Visual Display Division and Samsung Advanced Institute of Technology (SAIT) as a Research Master in the field of micro-optical systems with applications to imaging and display systems. From 2016, he joined KAIST as professor of NOVIC+ (Noise \& Vibration Control Plus) at the Department of Mechanical Engineering devoting to research on vibration, acoustics, vision sensors, and condition monitoring with AI.
His research fields include structural vibration; condition monitoring from sound and vibration using AI; health monitoring sensors; and 3D sensors, and lidar for vehicles and robots. He is the conference chair of MOEMS and miniaturized systems in SPIE Photonics West since 2013. He is a vice-president of KSME, KSNVE, KSPE, and member of IEEE and SPIE

\end{IEEEbiography}

\vfill

\end{document}